\documentclass[aps,nofootinbib,floatfix]{revtex4} 
\usepackage{graphicx}
\input epsf
\newcommand{\sfig}[2]{
\includegraphics[width=#2]{#1}
        }
\newcommand{\Sfig}[2]{
    \begin{figure}[thbp]
    \sfig{#1.eps}{.7\columnwidth}
    \caption{{\small #2}}
    \label{fig:#1}
    \end{figure}
}

\newcommand{\rf}[1]{\ref{fig:#1}}

\def\vl{{\bf l}}
\def\vtm{ $\boldmath$\theta$\unboldmath$ } 
\def\vt0m{ $\boldmath$\theta_0$\unboldmath$ } 

\def\cmm2{{\,\rm cm^{-2}}}
\def\cm2{{\,{\rm cm}^2}}
\def\cmm3{{\,{\rm cm}^{-3}}}
\def\gcmm3{{\,{\rm g\,cm^{-3}}}}

\def\fun#1#2{\lower3.6pt\vbox{\baselineskip0pt\lineskip.9pt
  \ialign{$\mathsurround=0pt#1\hfil##\hfil$\crcr#2\crcr\sim\crcr}}}

\def\be{\begin{equation}}
\def\ee{\end{equation}}
\def\bea{\begin{eqnarray}}
\def\eea{\end{eqnarray}}
\newcommand{\vs}{\nonumber\\}

\newcommand{\ec}[1]{Eq.~(\ref{eq:#1})}
\newcommand{\Ec}[1]{(\ref{eq:#1})}
\newcommand{\eql}[1]{\label{eq:#1}}

\begin{document}

\title{Reduced Shear Power Spectrum}

\author{Scott Dodelson$^{1,2,3}$, Charles Shapiro$^{4,5}$, Martin White$^{6,7}$}

\affiliation{$^1$NASA/Fermilab Astrophysics Center Fermi National
Accelerator Laboratory, Batavia, IL~~60510-0500}
\affiliation{$^2$Department of Astronomy \& Astrophysics, The
University of Chicago, Chicago, IL~~60637-1433}
\affiliation{$^3$Department of Physics and Astronomy, Northwestern
University, Evanston, IL~~60208} 
\affiliation{$^4$Department of Physics, The
University of Chicago, Chicago, IL~~60637-1433}
\affiliation{$^5$Kavli Institute for Cosmological Physics, The
University of Chicago, Chicago, IL~~60637-1433}
\affiliation{$^6$Department of Astronomy, University of California, Berkeley,
CA~~94720}
\affiliation{$^7$Department of Physics, University of California, Berkeley,
CA~~94720}

\date{\today}
\begin{abstract}
Measurements of ellipticities of background galaxies are sensitive to the
{\it reduced shear}, the cosmic shear divided by $(1-\kappa)$ where
$\kappa$ is the projected density field. We compute the difference between
shear and reduced shear both analytically and with simulations. The difference
becomes more important on smaller scales, and will impact cosmological parameter
estimation from upcoming experiments. A simple recipe is presented to carry out the required correction.
\end{abstract}
\maketitle

\section{Introduction}

One of the most fascinating aspects of general relativity, and the
first triumph for the theory, is that gravitational potentials can
act as lenses for light from distant sources.
The presence of large-scale structure and its associated potentials
along the line-of-sight to distant galaxies implies that the images
of most distant galaxies are slightly sheared compared to their
intrinsic shapes.  This shearing effect encodes information about
cosmological distances and the evolution of large-scale structure.
For this reason gravitational lensing of background galaxies by large
scale structure (cosmic shear) offers an excellent way to study the
distribution of matter in the universe
\cite{Kaiser:1991qi,Mellier:1998pk,Bartelmann:1999yn,Dodelson}. 
Measurements of the cosmic shear are already enabling us to constrain
the dark matter abundance and clustering 
amplitude among other 
parameters~\cite{Hoekstra:2002nf,VanWaerbeke:2004af,Heymans:2004zp,Jarvis:2005ck}.
In the future, large surveys may well uncover properties of dark energy,
such as its abundance and equation of 
state\,\cite{Benabed:2001dm,Refregier:2003xe,Takada:2003ef,Heavens:2003jx}, 
and of neutrinos\,\cite{Hu:1998az,Abazajian:2002ck}.
This program will be successful only if we can make very accurate theoretical
predictions \cite{Huterer:2004tr}.

As experiments begin to go deeper and cover more and more sky, theorists must
make sure that predictions are accurate enough to extract cosmological
information in an unbiased fashion. There is a quantitative way to phrase this
directive: the systematic errors on cosmological parameters induced
by theoretical uncertainties should be significantly smaller than the
anticipated statistical errors. Since the latter hover near the  
percent level for the most ambitious experiments, theorists clearly have their
work cut out for them.

Here we consider one correction to the standard theoretical predictions,
the effect of reduced shear \cite{Schneider:1997ge,Barber:2001yx,White:2005jr}.
The observed ellipticities of galaxies (with two components $g_i$ for
each galaxy) are often used as estimates of the cosmic shear ($\gamma_i$),
but in fact they are sensitive to the reduced shear: 
\be
  g_i={\gamma_i\over1-\kappa},  \quad \left| \kappa \right| < 1
\eql{defred}\ee
where $\kappa$ is the convergence or roughly the projected density field.
We analyze the difference between shear and reduced shear both analytically
and with numerical simulations \cite{Vale:2003ad,White:2003xz}, focusing on
various two-point functions. 
Then we map out the region in experiment-space where the effects of
reduced shear need to be included. 
Outside of this region, the canonical prediction -- which neglects
the $1-\kappa$ denominator -- is sufficient.

Throughout we assume a flat, $\Lambda$CDM cosmology.  Our base model has
$\Omega_m=0.28$, $\Omega_bh^2=0.024$, $h=0.7$, $n=1$, and $\sigma_8=0.9$.
We will let the galaxy density and sky coverage of surveys vary, but
we limit ourselves to all background sources at $z=1$. Our quantitative results will
change slightly for higher redshift sources, but our two major conclusions -- that
we can compute 
these corrections accurately and that we have to compute them if we want to extract cosmological
parameters -- are only strengthened (since the effect is larger) for higher redshift sources.

\section{Shear Two-Point Functions}

Here we briefly review a variety of definitions relating to cosmic shear
and its statistics. Cosmic shear can be represented by two
numbers at any point in space, $\gamma_1$ and $\gamma_2$. Similarly, the
ellipticity of a background galaxy can be described by $g_1$ and
$g_2$. The latter are measurable, while the former are related in
a straighforward way to the projected gravitational potential and therefore
are simplest to compute given a cosmological theory. On average, all these
components are zero; however their two-point functions contain information about
cosmic fluctuations.

We focus here on four sets of two-point functions.

\begin{itemize}

\item {\bf Smoothed Variance} Defining $\gamma\equiv
  \gamma_1+i\gamma_2$, make a map of $\gamma$ smoothed over a square
pixel of side $\theta$.  The variance of the smoothed shear field,
  $\bar\gamma$, is then
\be
\langle|\bar\gamma^2|\rangle(\theta) = \langle \bar\gamma_1^2\rangle(\theta) +
\langle\bar\gamma_2^2\rangle(\theta)
\ee
with a similar definition for the galaxy ellipticities, $g$.  Note
  that ``$\bar{\quad}$'' denotes an average over a local square while
  angle brackets denote an average over the sky.

\item {\bf Aperture Mass} At any sky position $\vt0m$, this is a 2D integral
over the {\it tangential shear\/},
$\gamma_t\equiv -\gamma_1\cos(2\phi) - \gamma_2\sin(2\phi)$,
where $\phi$ is the angle between $\vt0m$ and a fixed $x$-axis.
The weighting function in the integral depends on a smoothing scale $\theta$.
Here we use the smoothing function defined in \cite{Schneider:1997ge}.
The average value of $M_{\rm ap}$ is zero, but its variance as a function of
smoothing scale contains information about the underlying fluctuations.

\item {\bf Correlation Function} Since there are two components of shear, there
are in principle 3 separate correlation functions $\langle
\gamma_i(\vt0m)\gamma_j(\vt0m+\vtm)\rangle$ averaged over
all positions $\vt0m$. These depend on the angular difference
$\theta\equiv|\vtm|$. Here we focus on the combination
$\xi(\theta)\equiv\langle\gamma(\vt0m)\gamma^*(\vt0m+\vtm)\rangle$. 
\item {\bf Angular Power Spectrum} Write the $\gamma_i$ field as a sum of
coefficients times spherical harmonics. In the small angle limit in which we
will work, this is equivalent to a Fourier transform, $\tilde\gamma_i(\vl)$. One
linear combination of these Fourier coefficients (the so-called
``B-mode'') vanishes if the underlying fluctuations are due to scalar
perturbations; the other,
\be
\tilde E(\vl) = \epsilon_{ij} T_i(\vl) \tilde\gamma_j(\vl),
\eql{defe}
\ee
is sensitive to the projected gravitational potential. Here 
$\epsilon_{ij}$ is the 2D anti-symmetric tensor
$\epsilon_{12}=-\epsilon_{21}=1$; and the
trigonometric weighting functions are
\bea
T_1(\vl)&\equiv&-\sin(2\phi_l)
\vs
T_2(\vl)&\equiv&\cos(2\phi_l)
\eea
where $\phi_l$ is the angle of $\vl$ with a fixed $x$-axis.
The angular power spectrum is roughly the variance of these Fourier coefficents,
\be
\langle \tilde E(\vl) \tilde E(\vl') \rangle
= (2\pi)^2 \delta^2\left( \vl+\vl'\right) C_l
.\ee

\end{itemize}

Each of these two-point functions can be computed from a simulation
using either shear or reduced shear. Thus, for example, we can measure in a
simulation both $C_l^\gamma$ and $C_l^g$ and find the difference between the
two. The simulations we use to compute these functions are described in
Ref.~\cite{White:2005jr}.
We can also compute the two-point functions semi-analytically. The
two-point functions of $\gamma$ can be expressed in terms of integrals over
the 3D matter power spectrum, which has been extremely
well-studied~\cite{Hamilton:1991es,Peacock:1993xg,Peacock:1996ci,
Bernardeau:2001qr,Smith:2002dz}. The
two-point functions of $g$ can be computed perturbatively by expanding the
$1-\kappa$ denominator around $\kappa=0$. In \S III, we write the 
shear two-point functions as integrals of the 3D matter power spectrum
(these expressions are well-known~\cite{Kaiser:1991qi,Dodelson}) and
the reduced shear corrections in terms of the 3D three-point function,
the matter bispectrum.

\section{Perturbative Calculation}

The simplest two point function to compute is the angular power spectrum.
The Fourier transformed shear can be expressed in terms of the projected gravitational potential
\be
\tilde\gamma_i(\vl) = -\epsilon_{ij}T_j(\vl){l^2\over 2} \tilde\psi(\vl)
\eql{gampsi}\ee
with 
\be
\tilde\psi(\vl)\equiv \int_0^\infty {d\chi\over \chi} W(\chi) \int {dk_3\over 2\pi}
\tilde\Phi(\vl/\chi,k_3;\chi)
.\ee
Here $\chi$ is comoving distance; the lensing kernel
$W=2\chi(1-\chi/\chi_s)\Theta(\chi_s-\chi)$ with $\chi_s$ the distance to the
source galaxy and $\Theta$ the Heaviside step function. 

Inserting \ec{gampsi} into \ec{defe} to get $\tilde E(\vl)$, multiplying $\tilde
E(\vl)$ by $\tilde E(\vl')$, taking the expectation value, and then integrating
over $\vl'$ in the Limber approximation leads to
\be
C_l = {l^4\over 4} \int {d\chi W^2(\chi) \over \chi^6} P_\Phi(l/\chi;\chi)
\eql{clzero}\ee
where $P_\Phi$ is the 3D power spectrum of the gravitational potential.
The top panel of Fig.~\rf{clcomp} shows the power spectrum computed in this fashion
as compared with that measured in simulations. Agreement is excellent, confirming earlier
work~\cite{jainseljak,Hu:2000ax}.
The one aberrant point on small scales in the simulations is 
close to the Nyquist frequency, so power in the simulation is artifically
suppressed.

\Sfig{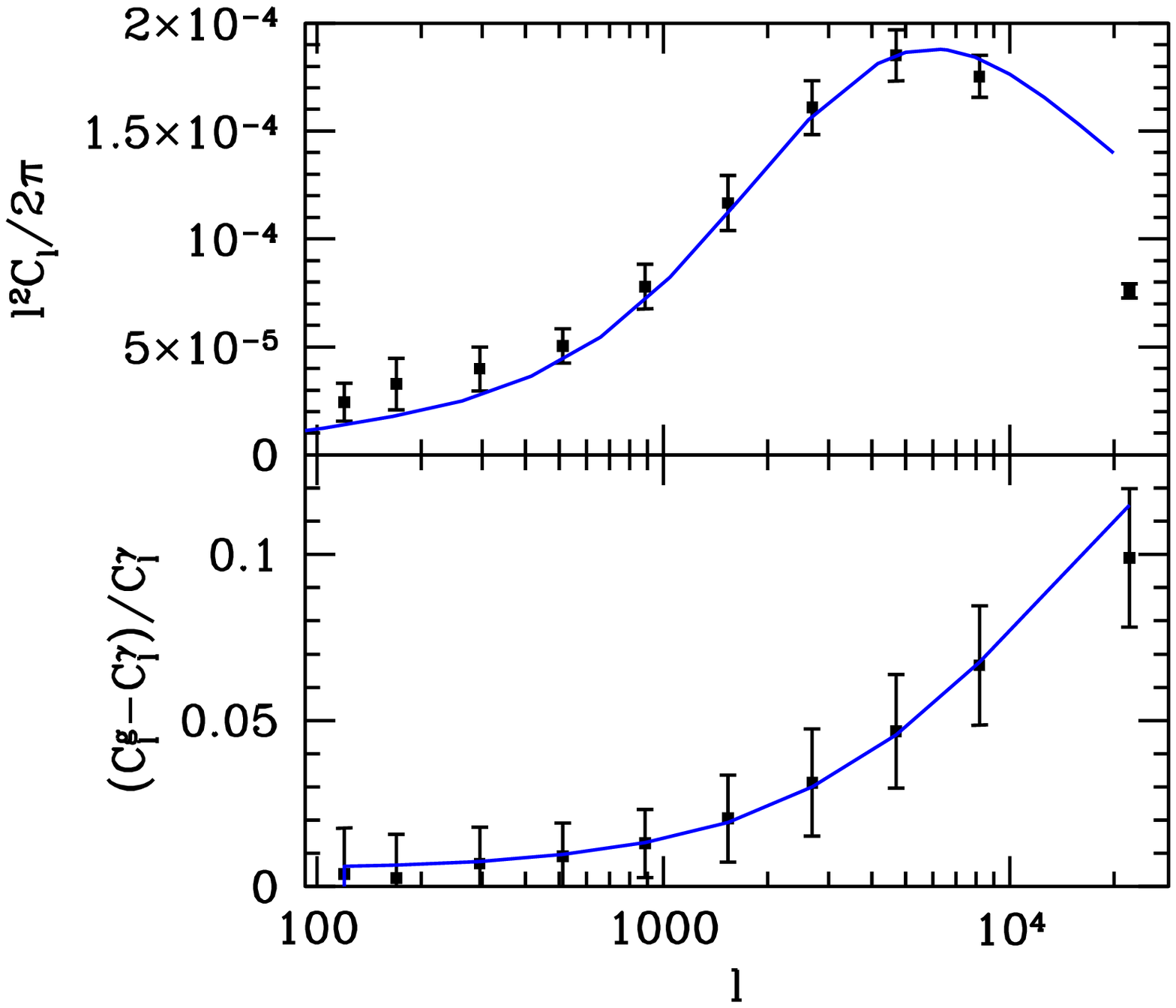}{Results from simulations (points with error bars) and
the perturbative calculation (smooth curves) for the zero order power
spectrum (top panel) and the fractional change due to reduced shear
(bottom panel).}

Eq.~\Ec{clzero} is an expression for the power spectrum of cosmic shear, $C_l^\gamma$. To lowest order,
when $\kappa\rightarrow 0$ in \ec{defred}, the power spectrum of the observable reduced shear is equal to this. 
We can perturbatively compute the correction to the reduced shear:
to leading
order, $g^{(1)}=\gamma$ and to second order $g^{(2)}=\gamma\kappa$. 
Therefore, the correction to the two-point function due to reduced shear is
\be
\delta \langle \tilde E(\vl) \tilde E(\vl') \rangle
= \epsilon_{ij}\epsilon_{kl} T_i(\vl) T_k(\vl')
\langle \tilde g_j^{(2)}(\vl) \tilde g_l^{(1)}(\vl') \rangle
+ \left( \vl\leftrightarrow \vl' \right)
\eql{deltae}\ee
Plugging in for $\tilde E$ and using $\epsilon_{ij}\epsilon_{jk}=-\delta_{ik}$, we have
\be
\delta \langle \tilde E(\vl) \tilde E(\vl') \rangle
= {-l^2\over 8} T_i(\vl) \int {d^2l_1\over (2\pi)^2} T_i(\vl_1)
l_1^2(\vl-\vl_1)^2 
\langle \tilde\psi(\vl_1) \tilde\psi(\vl-\vl_1) \tilde\psi(\vl') \rangle
+ \left( \vl\leftrightarrow \vl' \right)
\ee
Using Eqs.~(19,20) and (22) in Ref.~\cite{dodzha},
we can reduce this to
\be
\delta C_l={2T_i(l)\over l^4}\int {d^2l_1\over (2\pi)^2}
T_i(\vl_1) l_1^2 (\vl-\vl_1)^2 B^\kappa(\vl_1,\vl-\vl_1,-\vl)
\eql{deltacl}\ee
where $B^\kappa$ is the bispectrum of the convergence. Just as the power spectrum of the convergence can be written
as an integral of the 3D power spectrum along the line-of-sight (\ec{clzero}), the 2D bispectrum is an integral of
the 3D bispectrum~\cite{castro}:
\be
B^\kappa(\vl_1,\vl_2,\vl_3)= {-l^6\over 8}
  \int_0^\infty d\chi {W^3(\chi)\over \chi^{10}}
B_\Phi(\vl_1/\chi,\vl_2/\chi,\vl_3/\chi) .
\ee

The 3D bispectrum, $B_\Phi$, has been computed analytically on large scales
and measured on a wide range of scales in simulations~\cite{Bernardeau:2001qr}. 
An accurate fit to the N-body results was introduced in
Ref.~\cite{Scoccimarro:2000ee}; we use this fit to compute $\delta
C_l$, the difference between cosmic shear and reduced shear power.  The power
spectrum, $P_\Phi$, which is needed to compute $B_\Phi$, was
computed using the publicly available Halofit code \cite{Smith:2002dz}
which has also been calibrated by numerical simulations.
Fig.~\rf{clcomp} shows the results of this perturbative calculation and
of a similar measurement from simulations.
The perturbative results are in excellent agreement with the simulations.
This is extremely encouraging because it offers an easy way to include
reduced shear corrections without resorting to expensive simulations.

The conclusion that reduced shear differs from cosmic shear most significantly
on small scales follows from perturbation theory.
On large scales, fluctuations in $\gamma$ and $\kappa$ are small; since the difference
between cosmic shear and reduced shear is higher order in these perturbations
$(\propto \kappa\gamma)$, it is very small on large scales.

The angular two-point functions described in \S II can all be expressed as
integrals over the power spectrum. The smoothed variance is
\be
\langle|\bar\gamma^2|\rangle(\theta) =
  {1\over \pi^2} \int d^2l \tilde j_0(l_1\theta/2) j_0(l_2\theta/2) C_l
.\ee
This variance can be computed for either shear or reduced shear.
The difference between the two is the same integral over $\delta C_l$ from
\ec{deltacl}. The other spatial functions are
\be
\langle M_{\rm ap}^2\rangle(\theta) = {288 \over \pi\theta^4} \int_0^\infty dl
{C_l J_4^2(l\theta)\over l^3}
\ee
and
\be
\xi(\theta) = {1\over 2\pi} \int_0^\infty dl lJ_0(l\theta) C_l 
.\ee

\Sfig{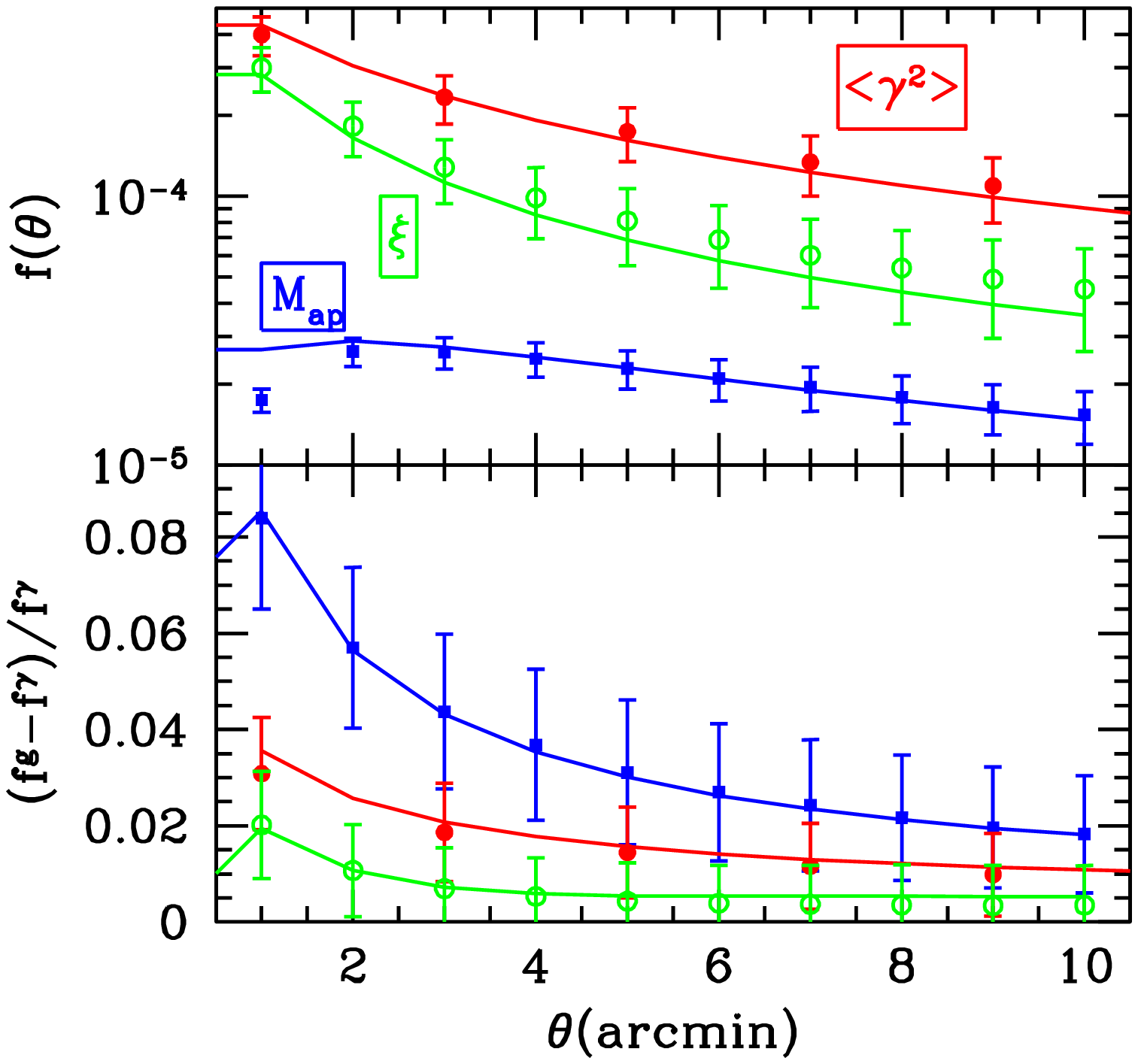}{Angular two-point functions from simulations and
perturbatively.}

We have computed these three functions from simulations and perturbatively;
the results are shown in Fig.~\rf{anglecomp}. The simulations and perturbative
calculations agree extremely well, as do the corrections.

\section{Impact on Cosmological Parameters}

When does one need to include the effects of reduced shear when comparing
models with observations? Neglecting these effects results in an incorrect
prediction for the power spectrum; we computed a correction, $\delta
C_l$, above. This incorrect prediction propagates to an incorrect
estimate of the  cosmological parameters, $p_i$, or a bias. The bias
on parameter $i$ is
\be
b_i \equiv p_i^{\rm true}-p_i= F^{-1}_{ij} \sum_l w_l {\partial C_l\over \partial p_j}
\delta C_l
.\eql{bias}\ee
Here $w_l$ is the weight, the inverse variance of the measurement, and $F$ is
the Fisher matrix
\be
F_{ij} =\sum_l w_l {\partial C_l\over \partial p_i} {\partial C_l\over \partial
p_j}
.\ee
The variance depends on experimental specifications: sky coverage and depth/resolution. Specifically,
\be
w_l^{-1} = {2\over (2l+1) f_{\rm sky}} \left( C_l + {\langle \gamma_{\rm int}^2 \rangle \over n_{\rm eff}} \right)
.\eql{wl}\ee
The fraction of sky covered is $f_{\rm sky}$, while the rms of the intrinsic ellipticity of galaxies,
$\langle\gamma_{\rm int}^2\rangle^{1/2}$, is set to
$0.25$~\cite{Hoekstra:1999pu,Refregier:2003xe}, and $n_{\rm eff}$ is the effective
galaxy density which depends mainly on the
depth and resolution of the experiment.

We can compute the bias induced by neglecting the difference between shear
and cosmic shear for any experiment. For concreteness, we allow three
cosmological parameters to vary: the normalization of the power spectrum,
$\sigma_8$; the shape of the power spectrum, $\Gamma\simeq\Omega_mh$;
and the matter density, $\Omega_m$.
Figure~\rf{bias} shows the bias induced in these parameters by neglected
the cosmic shear corrections. This bias was computed including data out
to $l=3000$. The cut-off is necessary because baryons affect the theoretical
predictions on small scales \cite{White:2004kv,Zhan:2004wq}, and it is very
difficult to predict these effects accurately.

\Sfig{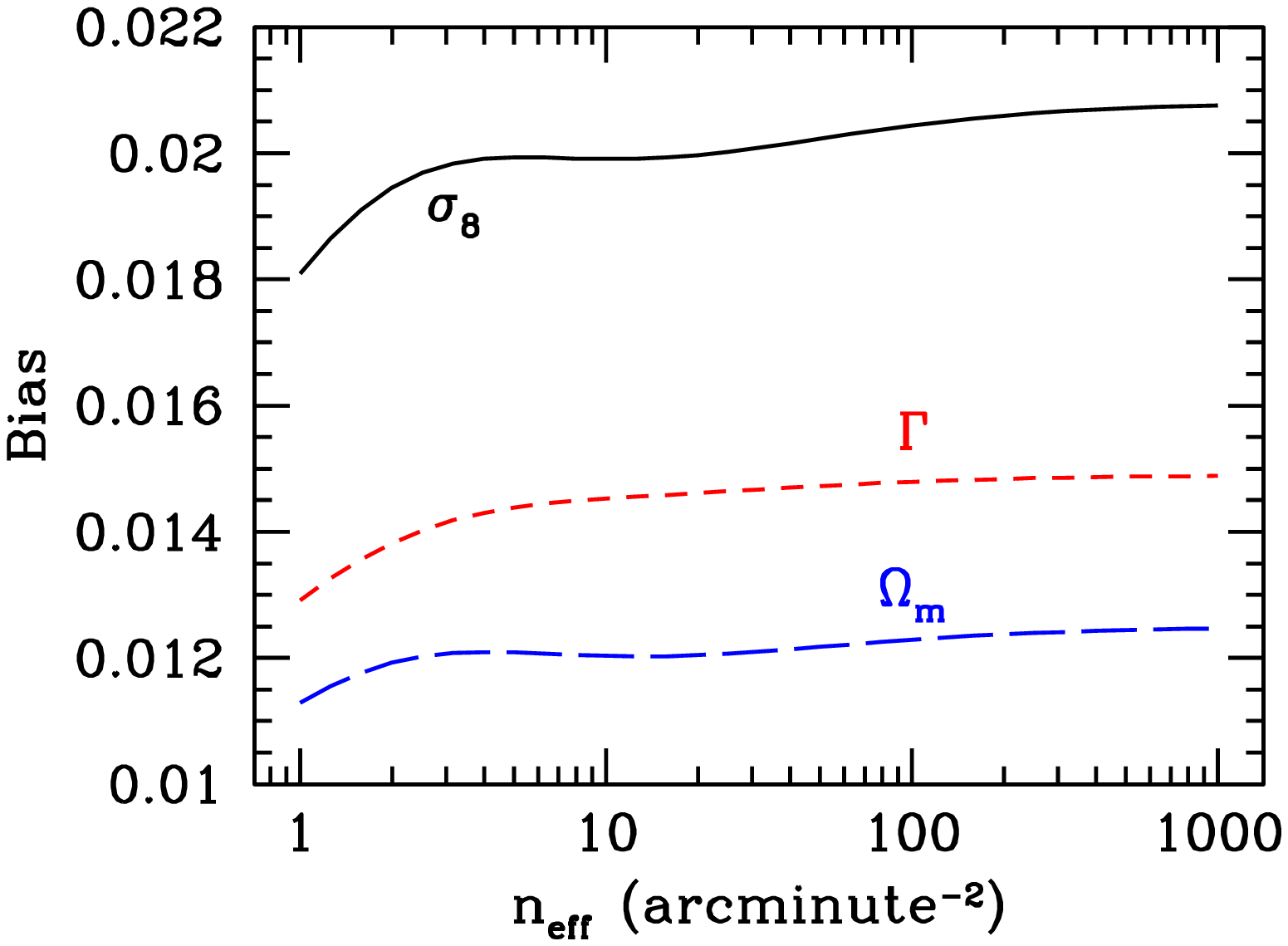}{The bias due to neglecting reduced shear corrections as a
function of galaxy density (per square arcminute for a 1-component rms
shear of $0.25$) for the three cosmological parameters varied.
The bias on $\sigma_8$ is negative.}

Since the reduced shear corrections on scales $l<3000$ are of order a few
percent (Fig.~\rf{clcomp}), it is not surprising that the induced biases
on the parameters are also of order a few percent.
Note from \ec{bias} that the bias scales as $F^{-1} w_l$.
Since  $w_l\propto f_{\rm sky}^{-1}$, and $F^{-1}$ scales as $f_{\rm sky}$
the bias is nearly independent of sky coverage.
Therefore, the bias depends only on the galaxy density.
At very large density, the weight $w_l$ becomes independent of galaxy density,
since shape noise due to intrinsic ellipticity is inversely proportional to
galaxy density.
In this limit, cosmic variance -- the first term in parentheses in \ec{wl} --
becomes the dominant source of noise.  The largest bias is to the
normalization parameter $\sigma_8$. 

How important is a one to two percent level bias in the cosmological
parameters? We must compare the bias to the anticipated statistical
error in an experiment. If there were only one free parameter, this
comparison would be straightforward. With several parameters, we must
ask which statistical error should be used as a baseline: the error on
$\sigma_8$ for example which accounts for the uncertainty in all other
parameters (this is called the {\it marginalized\/} error) or the
error expected if all other parameters are held fixed?  We argue that
the latter should be used. To understand why, consider the case with
two parameters. The expected 1-$\sigma$ constraints trace out an
ellipse in this plane. If the parameters are degenerate\footnote{A
  smart analyst tries to choose non-degenerate parameters; in that
  case, the marginalized and   unmarginalized errors are equal.}, this
ellipse is very elongated, and the expected constraints on either of
the parameters individually will be very weak. If these marginalized
errors are used, then the ratio (bias/error) will be quite small. A
bias much smaller than the marginalized statistical error, however,
can easily induce an analyst to estimate the parameter as lying
outside the allowed ellipse. If we instead use unmarginalized errors
as the baseline, then bias/statistical error ratios smaller than one are
more likely to keep the parameters within the allowed ellipse.
Fig.~\rf{s8ratioum} shows the ensuing ratio of bias to statistical
error for the most severely affected cosmological parameter
($\sigma_8$) as a function of survey width ($f_{\rm sky}$) and galaxy
density.

\Sfig{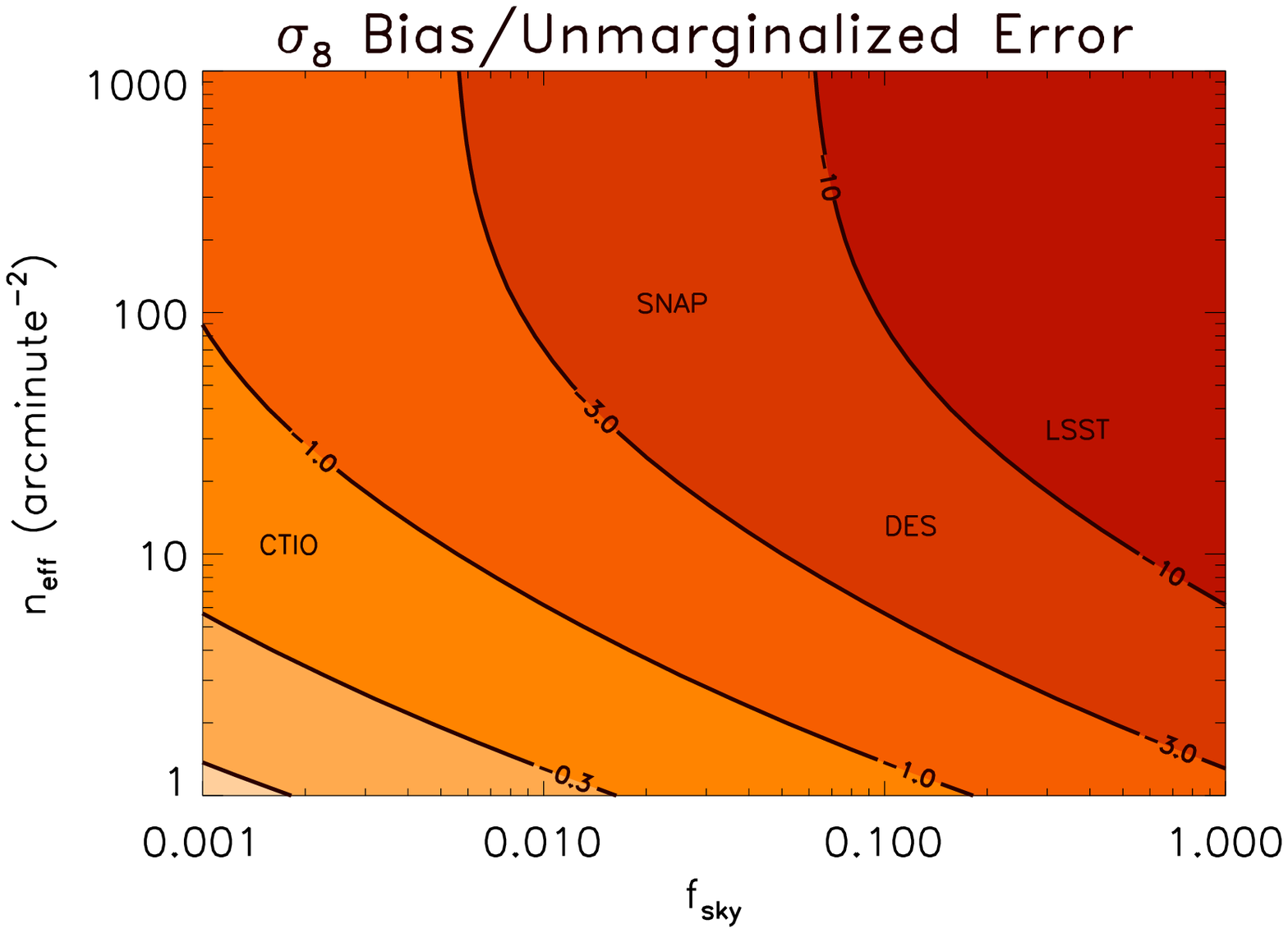}{The ratio of the bias on $\sigma_8$ due to neglecting reduced shear
to the statistical error on $\sigma_8$. The ratio depends on the width and galaxy density
of the survey. Several current and future surveys are shown for orientiation: CTIO Lensing
Survey~\cite{Jarvis:2005ck}; Supernova Acceleration
Probe (SNAP)~\cite{snapwl}; Dark Energy Survey (DES)~[{\tt http://www.darkenergysurvey.org}]; 
Large Scale Synaptic Survey (LSST)~\cite{Tyson:2002nh}. 
If the ratio is larger than
one, neglecting reduced shear corrections biases the parameters by an amount greater than
the unmarginalized statistical errors.}

Alternatively, we can transform to a basis in parameter space where the
Fisher matrix is diagonal and then rescale so that the errors on the
parameters are unity. In this basis, magnitude of the bias is a simple measure
of how far outside (or inside) the 1-sigma contours the bias leads. Again
a bias greater than unity means the effect needs to be accounted for.
Figure~\rf{bcontour} shows the bias in this rotated/rescaled basis. 
For example, neglecting the bias in the dark energy survey leads to an incorrect estimate
of parameters at the 3-sigma level.

\Sfig{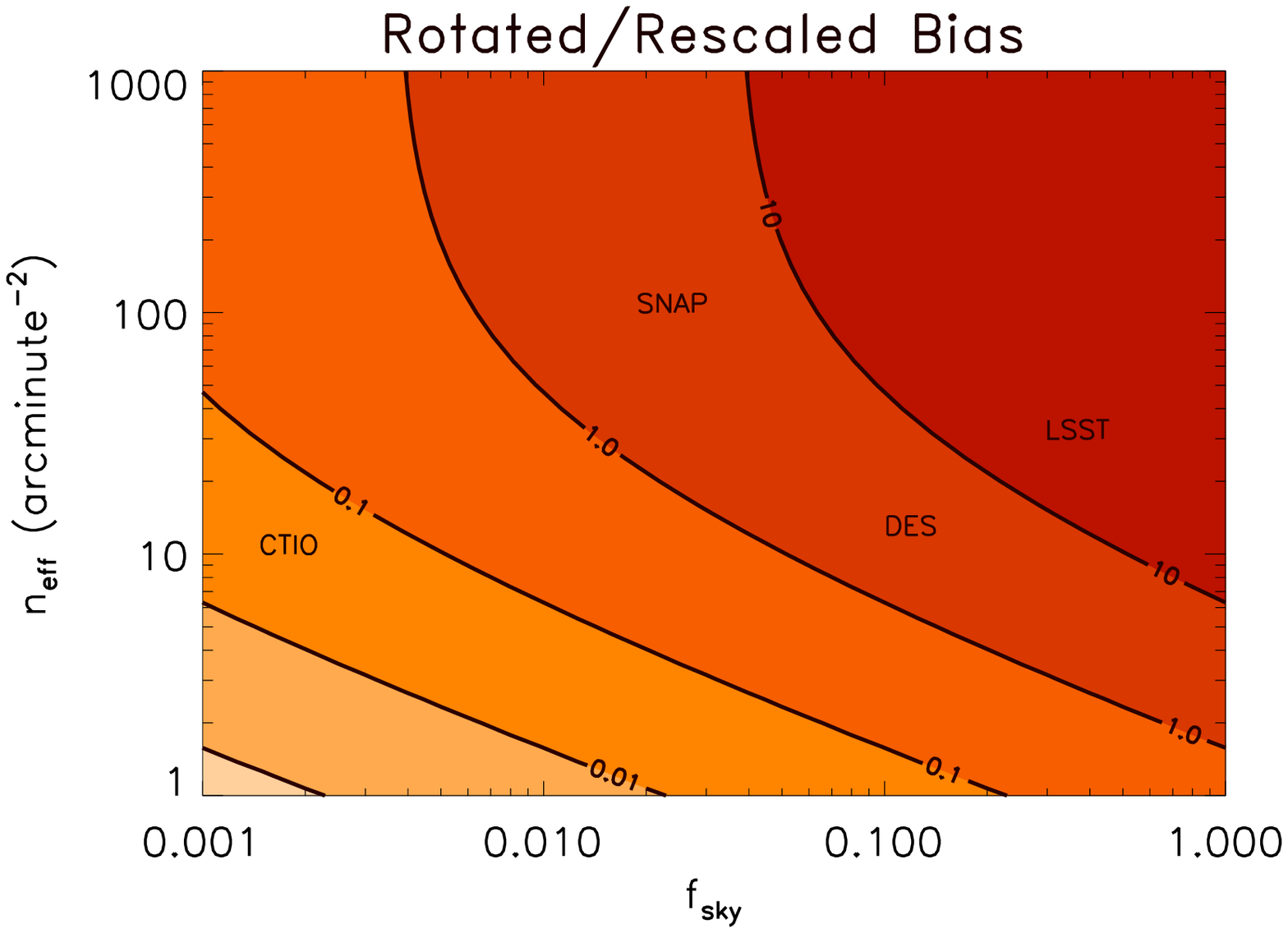}{The bias for rotated and rescaled parameters such that the anticipated
errors are diagonal and unity. That is, the 1-sigma contours in the 3D parameter space map out 
a sphere of radius one in this basis. The amplitude of the bias is shown here as a function
of galaxy density and sky coverage.}

Both Figures \rf{s8ratioum} and \rf{bcontour}
illustrate that the largest of current experiments are already
entering a regime in which the effects of reduced shear must be
considered and future experiments will certainly need to account for it.
Relying on Fig.~\rf{bcontour} to justify neglecting the effect for current experiments
is a little dangerous since it does not account for priors, which can reduce errors
while leaving the bias unaffected. So the most conservative approach is to use Fig.~\rf{s8ratioum}
as a guide to the impact of reduced shear. In that case, virtually all upcoming experiments
will need to add in this correction when extracting information about cosmological parameters.

\section{Other Corrections}

Reducing the shear is not the only correction that needs to be applied to
weak lensing spectra~\cite{jainseljak,Schneider:1997ge,Vale:2003ad}.  For instance, Hu and Cooray~\cite{Cooray:2002mj}
have computed perturbative corrections to the power spectrum which
account for the Born approximation and lens-lens coupling.  These corrections
are an order of magnitude smaller than the one considered here. To understand why,
recall that $\delta C_l$ from \ec{deltae} comes from considering the product of
a second order perturbation with a first order perturbation, $g^{(1)} \times g^{(2)}$.
We have considered the  corresponding terms for the beyond Born and lens-lens corrections
and found that they vanish. Therefore, the first non-vanishing corrections are the ones
considered in Ref.~\cite{Cooray:2002mj}: those of order $g^{(1)} g^{(3)}$ or $g^{(2)}g^{(2)}$.
We have also not considered the effects of source clustering~\cite{Bernardeau:1997tj}.
There is some indication~\cite{Hamana:2001kd,2002A&A...389..729S} that this may also induce percent level changes in
the power spectrum on small scales. If so, these would need to be included as well. Accounting for 
this coupling in a simulation requires input from a real 3D galaxy catalogue.

\section{Conclusions}

Deflection of light rays by gravitational potentials along the line of
sight introduces a mapping between the source and image plane.  The
Jacobian of this mapping defines the shear and convergence as a function
of position on the sky.  In the absence of size or magnification information
neither the shear nor the convergence is observable, but only the combination
$g=\gamma/(1-\kappa)$, known as the reduced shear.
On small scales, where lensing surveys get much of their constraining power,
this must be taken into account when predicting the observables.

We have studied the difference between cosmic shear and the reduced shear.
Our main conlusions are:
\begin{itemize}
\item The perturbative calculation of reduced shear, \ec{deltacl}, agrees
well with numerical simulations.  This is not too surprising, since the
lensing is weak and we use many components themselves fit to N-body
simulations, but it gives us confidence that we can use these calculations
in making predictions or fitting to data.

\item The effects of reduced shear are on the threshhold of becoming very
relevant. As depicted in Figs.~\rf{s8ratioum} and \rf{bcontour}, upcoming surveys will need
to account for reduced shear when extracting cosmological parameters.
\end{itemize}

SD was supported by the DOE and by NASA grant NAG5-10842.
MJW was supported in part by NASA and the NSF.
CAS was supported in part by the Kavli Institute for Cosmological
Physics at the University of Chicago and by grant NSF PHY-011442.
The simulations were performed on the IBM-SP at NERSC.

\bibliography{cl3}
\end{document}